\documentstyle[12pt,twoside,cmp209,epsfig]{article}

\newcommand{\be}{\begin{equation}}
\newcommand{\ee}{\end{equation}}

\title[Foundations of Statistical Mechanics]
	{Foundations of Statistical Mechanics: in and out of Equilibrium}
\author[D. Karevski]
	{D. Karevski}
\address{
         {Laboratoire de Physique des Mat\'eriaux, }
        {UMR CNRS 7556,}\\
        {Universit\'e Henri Poincar\'e,  Nancy 1,}\\
        {F-54506 Vand\oe uvre les Nancy Cedex, France}\\
}

\begin{document}

\maketitle

\begin{abstract}
The first part of the paper is devoted to the foundations, that is the mathematical and physical justification, of  equilibrium statistical mechanics. It is a pedagogical attempt, mostly based on Khinchin's presentation, which purpose is to clarify some aspects of the development of statistical mechanics.
In the second part, we discuss some recent developments that appeared out of equilibrium, such as fluctuation theorem and Jarzynski equality. 
\ { Classical ensemble theory  }
	{05.30.Ch}\ { Quantum ensemble theory  }
	{05.70.Ln}\ { Non-equilibrium and irreversible thermodynamics}
\end{abstract}

\section{Introduction}
The main goal of statistical mechanics, at least from the point of view of the initiators like Boltzmann, Maxwell, Gibbs\footnote{Gibbs had a pragmatic point of view which was somehow different from Boltzmann's view. Gibbs founded statistical mechanics as a branch of rational mechanics, no matter what physical process generates the distribution in phase space. Contrary to that, Boltzmann's view-point was to really reduce thermodynamics to mechanics and consequently he necessitated an explanation of the mechanism that lead to equilibrium a mechanical system which was  initially in a nonequilibrium state. To skech the differences one can say that the Boltzmann approach is more physical whereas the Gibbsian is more rigourous.}, 
Einstein, was the derivation of the thermodynamical laws from the microscopic (atomistic) structure of matter. All these attempts are subjected to start from some models of the structure of matter. But it is a well known fact that thermodynamics was constructed independently or at least following a parallel road on the basis of few foundamental laws that are viewed as empirical facts. Very recently, Lieb and Yngvason tried to clarify some aspects of the second law (entropy) of thermodynamics on the basis of the concept of adiabatic accessibility. This work was mainly motivated by the fact that usual formulations of the second law, such as Kelvin or Clausius, use concepts such as hot, cold or heat that are intuitive but not really well defined nor precise before the theory is fully developed. Their basic derivation of the second law (that is the existence of the entropy state function) is based on some abstract postulates of a certain kind of ordering on a set of states.

From our point of view, the problem of the foundations of statistical mechanics is two-fold. One is: given a statistical theory, one has to extract quantities (averages of phase functions) and laws that can be identified with thermodynamical quantities and fundamental laws. The identification itself being of analogy type.\footnote{As it is very explicitly emphasised in Gibbs treaty\cite{gibbs}} Once those fundamental laws are recognized, one can logically develop the entire consequences of these laws. This logical enterprise was perfectly achieved by Gibbs in his celebrated treaty\cite{gibbs}. 
The other, less easy task, is to justify the use of the statistical 
theory (precepts) itself from a realistic\footnote{Given that the basic ontology is a single mechanical system composed of many subsystems (particles).} point of view, that is, so to speak
to justify the very use of ensembles. Differently stated, 
why canonical or microcanonical ensembles are suitable to describe real physical systems ?
This question arises since it is generally believed that the system to be studied is in a definite state and not distributed over a continuum of states.
It is clear that this second task is more physically related to the very structure of matter, and it is  within this perspective, that the work of Boltzmann has to be viewed. 
The (partial) answer to this question is related to the fact that real thermodynamical systems are constituted, at least approximately, of a huge collection of particles. The discussion of these points will be largely developed in the next section. Section~3 deals with non-equilibrium aspects and we present their relations, such as the fluctuation theorem and Jarzynski equality, which seem to many physicists to be of fundamental interest. In the last section we present some results obtained on the Ising model in the fluctuation relation context.

\section{Foundations of statistical mechanics}
\subsection{Interpretation of physical quantities}
The state of our system (classically a point in the phase space or a Hilbert spectral ray quantum mechanically) fully determines the physical (dynamical) quantities which caracterize the given system. We will generally call such a quantity a phase function (classical case $f(q,p)$, quantum case $f(\psi)=(\psi,{\bf Q}\psi)$ where $\bf Q$ is the operator associated to the quantity $f$).
In order to have a suitable theoretical description, one has to identify such phase functions with the various physical quantities obtained experimentally from measurement processes and compare their respective value. However, in order to compare the empirical data with the theoretical predictions, one has to know the actual state of the system, that is, for example classically, to determine $2s$~($\sim 10^{23}$) coordinates. But in general, the empirical (macroscopic) description of what is called a (equilibrium) thermodynamical state is fully specified by a very small set of independant variables, such as the energy, volume, pressure, ...
So that the question that rises is which state should we choose in order to evaluate the relevant phase functions and compare their value with their experimental counterpart ? and obviously no one has or can have any reasonable answer to such a question. 
Nevertheless, if one realizes that the measurement of a physical quantity is performed during a finite time, which in general is very large compared to some internal time scale, one realizes that the actual empirical datas are given as averages of the quantities over long time periods. But the initial question still survives, that is, which (part of the)\footnote{One problem that arises is the fact that time average of a phase function on a given trajectory may have very different values for different time intervals. This difficulty is overcome thanks to a theorem due to Birkhoff, which states that for almost all trajectories, the time averages of a given phase function tend to a definite limit when the time interval tends to infinity. It means in particular that the averages over finite time intervals on a given trajectory (a typical one) will take approximately the same value for sufficiently large time-periods. This remark, basically, is at the
heart of the time average procedure used widely to start an exposition of statistical mechanics, see for example ref.\cite{landau}. } 
trajectory the system is actually following ?  In order to answer such a question one has to know $2s-1$ independant integrals of motion and it seems that a very small path has been done toward the solution since our starting puzzling problem of finding the $2s$ coordinates. 
At this step, as it is well known, in order to avoid the average over an unknown trajectory, normally one invoques ergodic theorems or hypothesis to replace time averages by phase averages. However, in general very few systems are known to be ergodic and it seems really unprobable that in a realistic case one will ever prove ergodicity. But the requirement of ergodicity is too strong. For instance one has simply to require that only few (corresponding to the empirical ones) phase functions should have time average equal to their phase average. So, it would be an unnecessary hypothesis to demand the validity of such an equality for all phase functions. Another objection that has to be emphasized is the fact that ergodicity is a requirement that involves average over recurrence time, which is too long to have any physical relevance (several astronomical orders of magnitude), but in actual experiment, the times involved to obtain the averages are by far shorter than the reccurence time, see for example for a discussion of this point Ref.\cite{gallavotti}. 
The reason of this is lying in the fact that the majority of phase functions describing physical quantities exhibit a very peculiar behaviour. They are approximately constant on almost all the points of the constant energy manifold (since we are talking here of an isolated system).  Why is it so is linked to the fact that the mechanical systems, considered here, are breaking up into a large number of components and the fact that the interesting phase functions are sum functions, that is sum of functions depending on the dynamical coordinates of the component subsystems alone. 

If we suppose, for some reasons, that we can replace time-averages by phase-averages, then the remaining problem is to determine the suitable phase average procedure. In the case of an isolated mechanical system, 
it is usually argued that one has to restrict the phase average to the constant energy manifold since the actual trajectory is taking place into this subset of the phase space. Indeed, if one consider the energy phase function for such an isolated system and one takes its phase average over several constant energy manifolds, it is clear that the average value will not give the actual energy value. To overcome this discrepency, one has to restrict the phase average on that constant energy subset corresponding to the real value. The argument is clear and easy to conceive. But one has to realize that every thing that has been said concerning the energy should also be true for the $2s-2$ other integrals of motion. So in particular, if we consider some other conserved quantity to be given as the energy is given, we have to restrict the average procedure to the intersection of the two corresponding manifolds, where the actual trajectory is taking place. Continuing on this lines, we will finally arrive at the fact that we have to average over the intersection of the $2s-1$ conserved quantities manifolds, that is over the actual trajectory and this is precisely what we wanted to avoid. In order to escape this vicious circle, let us consider a certain integral of motion defined by a phase function $I$.  If $I$ over the constant energy manifold takes approximately always the same value, then its time average over almost all trajectories will give almost the same (physical) value. Consequently, its phase average on the constant energy manifold 
will give a definite value which can be compared to the actual physical value. 
On the other hand, if this is not true and the phase function $I$ varies widely on the constant energy manifold, no definite average for different trajectories can be affected to it and it will not have a 
(macroscopic) physical interpretation.  However, if such a phase function has indeed a physical interpretation (and so an actual possibility of measuring it), one has to treat it on exactly the same footing as the energy is treated, and consequently if the value of that integral is known, we should restrict the phase average to the corresponding manifold of constant energy and constant $I$. It is a fact that in ordinary macroscopic systems, usually only the energy integral has to be considered.\footnote{One can refer to \cite{landau} for some more arguments relating the question to additive integrals of motion, which are explicitely, the energy, the momentum and angular momentum components, that is integrals that are related to the space-time symmetries. If one encloses the system into a rigid vessel at rest, translation and rotation invariance are broken and only the energy integral remains. The additivity argument relies on the fact that the equilibrium statistical distribution over the restricted phase space should have a product structure related to the statistical independence of the different components of the system, that is related to the small interaction between the different components.}

\subsection{Microcanonical principle}
As we have seen from the arguments given above, one of the goal of a suitable statistical theory is to give for some phase functions the same values as those obtained experimentally and compatible with phenomenological thermodynamics.
But, as it was argued, the values of the phase functions over a subset of the phase space present generally fluctuations that could be very large. In the following we give some more arguments in the direction of a 
phase space average. 

A thermodynamical state is specified completly by a
small set of quantities
\begin{equation}
{\cal Q}_1, {\cal Q}_2, ..., {\cal Q}_k\; .
\end{equation}

This means in particular that all physical quantities $B$ are given as functions of these variables:
\begin{equation}
B=f_B({\cal Q}_1, {\cal Q}_2, ..., {\cal Q}_k)\; .
\end{equation}
To be more precise, there is one such function, named the fundamental relation:
\begin{equation}
{\cal F}=f_{\cal F}({\cal Q}_1, {\cal Q}_2, ..., {\cal Q}_k)
\end{equation}
from which all the other quantities can be obtained by appropriate derivations. This is basically thermodynamics, at least if one specify some properties of that fundamental function (second principle).\cite{callen} 
On the microscopic side, it is clear that the specification of this very small set of quantities is not at all sufficient to completely determine, from the microscopic point of view, the state of the system. In general, for a given set $({\cal Q}_1, {\cal Q}_2, ..., {\cal Q}_k)$ we have many compatible microscopic states (let us call them as it is usual microstate). To be more specific, we will continue the discussion in the quantum case. So for the given set $({\cal Q}_1, {\cal Q}_2, ..., {\cal Q}_k)$ we have some set\footnote{Usually the set has the power of the continuum, so that the integer indexation used here should no be taken too seriously.}
$$
\psi_1,\psi_2,...,\psi_i,...
$$
of microstates to which are associated the corresponding values
$$
B_1,B_2,...,B_i,...
$$
of the phase function such that $B_i=f_B(\psi_i)$.

Let us imagine that we had prepared a collection of $N$ copies partially specified by the set $\cal Q$.
Measuring on each copy the quantity $B$ we obtain with some frequency $n_i/N$ the corresponding value $B_i$.
If we want to compare some theoretical value associated to the macrostate specified by the set $\cal Q$, we have to consider the average of the quantity $B$ over our experimental data. This is given by 
\begin{equation}
\langle B\rangle =\frac{1}{N}\left(n_1B_1+n_2B_2+...+n_iB_i+...\right)\; .
\end{equation}
The values $B_i$ are in principle known since they are quantum mechanical expectations over the states $\psi_i$.  The real problem comes from the fact that no theory can give the values of the frequencies $n_i/N$
since they are related to the actual experimental setup. Different devices, preparation protocols in the experimental setup will lead to different occurences of the microstates and so of the values of $B$. 
To solve the difficulty one could think of stating an average principle, that is of specifying the set $\{n_i/N\}$. But again, in general different average principles will lead to different $\langle B\rangle$.
In order to reconcile this observation with the uniqueness of empirical observations, one is naturally lead to the fact that almost all values $B_i$ should take almost the same value:
$$
B_i\approx \langle B\rangle = B^{Ther}
$$
The measure of the set where this equality doesn't hold should vanish or be reasonnably small. 
By calling upon the argument of simplicity, in order to calculate the average $\langle B\rangle$, we can choose the microcanonical average principle, that is specifying equal weights to all microstates.
One has to realize here that from the given argument the microcanonical average principle is not unique. But, nevertheless, one can argue that, given all the previous (more or less of euristic type) arguments and the fact that the microcanonical principle generates the same macroscopic relations as thermodynamics does, it is legitimate to postulate it by the Laplacien ``principle of insufficient reason''.\footnote{The principle of insufficient reason or principle of indifference, states that if there is no known reason for predicating of our subject one rather than another of several alternatives, then relatively to such knowledge the assertions of each of these alternatives have an equal probability.} 
According to an other guideline\cite{jaynes}, one can try to base the foundations on the ground set by information theory, that is to have a constructive criterion (maximum-entropy principle) for selecting a probability distribution on the basis of partial knowledge. In the case of constant energy systems, 
the maximum-entropy principle leads to the microcanonical distribution. 
To end this discussion, one can state again that not all phase functions should satisfy the average principle requirements, only those having a macroscopic thermodynamical interpretation. Indeed, no one is expecting that the averaged (microcanonical or whatever) one particle velocity of a gaz should give the actual measured velocity of a particule of that gaz. 

Let us consider for an isolate system the set of all the quantum states for which the energy is precisely fixed to the value $E$. These states are stationnary states, that is eigenstates of the Hamiltonian operator $H$ with eigenvalue $E$. The dimension of this Hilbert sub-space ${\cal H}_E$ is given by the degeneracy of the eigenvalue $E$:
\begin{equation}
\dim {\cal H}_E=\Omega(E)=m\; .
\end{equation}
Let $\psi_1,...,\psi_m$ be a complet orthonormal set of eigenvectors of $H$ spanning the entire subset ${\cal H}_E$. One has for all $\Psi\in {\cal H}_E$ the unique decomposition
\begin{equation}
\Psi=\sum_{i=1}^{m} \alpha_i \psi_i
\end{equation}
with the scalar coefficients $\alpha$ sitting on the complex hypersphere $S^*$
\begin{equation}
\sum_{i=1}^{m}|\alpha_i|^2=1\; 
\end{equation}
The microcanonical average of the phase function $f_Q(\Psi)=(\Psi,{\bf Q}\Psi)$ is given by the integral over the complex hypersphere with uniforme measure. One has
\begin{equation}
\langle Q\rangle \equiv \langle f_Q\rangle = \int_{S^*}(\Psi,{\bf Q}\Psi)dS^*\; .
\end{equation}
Together with the decomposition on the base $\psi_1,...,\psi_m$ and with the parametrization $\alpha_k=r_k e^{i\varphi_k}$, one arrives at
\begin{equation}
\langle Q\rangle = \sum_{k,l}(\psi_l,{\bf Q}\psi_k)\int_{S^*}\alpha_k\alpha_l^*dS^*
\end{equation}
\begin{equation}
\langle Q\rangle = \sum_{k,l}(\psi_l,{\bf Q}\psi_k)\frac{1}{m}\delta_{k,l}\; ,
\end{equation}
so that finally
\begin{equation}
\langle Q\rangle = \frac{1}{\Omega(E)}\sum_{k}(\psi_k,{\bf Q}\psi_k)=Tr_{{\cal H}_E} \{\frac{1}{\Omega(E)}{\bf Q}\}\; .
\end{equation}
Due to the invariance of the trace with respect to the change of orthonormal bases, the average  
$\langle Q\rangle$ is also invariant. This final rule is the basic starting point of a microcanonical calculation.\footnote{It is known that the state vectors should statisfy symmetry or antisymmetry principle
for bosons and fermions. This imply that in the given microcanonical average, one has just to restrict the subspace ${\cal H}_E$ to either symmetric subspace ${\cal H}^s_E$ or antisymmetric subspace ${\cal H}^{as}_E$}

\subsection{Suitability of the microcanonical principle}
In order that the microcanonical average of a quantity $Q$ has a physical signification, it is necessary that it takes a value close to the real experimental value or to the value given by a real dynamical theory. 
As it would be shown bellow, this requirement would be fulfilled by sum functions for which the average $\langle Q\rangle$ is of the order of the number of components $N$ of the system, that is
\begin{equation}
\langle Q\rangle =Nq
\end{equation}
where $q$ referes to the one component quantity (a density like value). 
To show the proposition, let us consider the measure $M_\delta$ of the set of the vector states $\Psi\in {\cal H}_E$ such that the probability 
\begin{equation}
P_{\Psi}(|Q-Nq|>N\epsilon)>\delta
\end{equation}
that the quantity $Q$ differ significantly (of the order $\epsilon N$) from the average value $Nq$ is greater than $\delta>0$. 
We have
\begin{equation}
\int_{S^*}P_{\Psi}(|Q-Nq|>N\epsilon) dS^* \ge \delta M_{\delta}
\end{equation}
from which, together with the Chebyshev inequality $\langle x^2\rangle\ge \epsilon^2 P(|x|>\epsilon)$, we arrive at
\begin{equation}
M_{\delta}\le \frac{1}{\delta \epsilon^2 N^2}\int_{S^*}
(\Psi,\left[{\bf Q}-Nq\right]^2\Psi) dS^*
\end{equation}
that is with the previous result on microcanonical average
\begin{equation}
M_{\delta}\le \frac{1}{\delta \epsilon^2 N^2} Tr_{{\cal H}_E} \{\frac{1}{\Omega(E)}{[\bf Q}-Nq]^2\}
=\frac{1}{\delta \epsilon^2 N^2} D(Q)\; ,
\end{equation}
where $D(Q)$ is the microcanonical dispersion of the quantity $Q$. 
Since in usual physical situations the dispersion is growing linearly (which is the content of the law of large numbers) with the number of components $N$, one arrives at 
\begin{equation}
\frac{D(Q)}{N^2}\sim \frac{1}{N}
\end{equation}
so that in the thermodynamical limit this ratio vanishes. This means that the quantity $Q$ would approximately agree with the phase average with a probability arbitrarily close to unity and this finally demonstrates the suitability of the microcanonical phase space average. We see here that the main reason for the validity of such a principle is the fact that thermodynamical systems are build up from myriades of component subsystems and that actual relevant physical quantities are sum functions.\cite{khinchin2}
Moreover, it is possible to show that the result just obtained remains valid if one affects some arbitrary absolutely continuous probability law, with respect to the measure introduced on the constant energy manifold, to the occurence of the state $\psi$.

\subsection{Canonical distribution}
Up to now, we have always considered isolated mechanical systems, but it is clear that such an idealization is not realistic nor efficient from a technical point of view. Indeed, a real system is never completely isolated and continuously interacts with its surounding. If the system is initialy prepared in a given (quantum) state, due to the interaction with its environment, it will very rapidly make transitions among its accessible states. Since such transitions are induced by purely random processes, for a sufficient time (it is claimed that) the system will sample out all the permissible states with equal probability.\cite{callen}

An other idealization is one in which the system under interest is free to exchange an arbitrary amount of energy with a very huge surrounding, the total system plus surounding being isolated. If the surrounding has good enough properties (it is then called a thermal bath), it basically fixes the temperature of the small (in comparison with the bath) system.  It is possible to show rigourously that if we accept the principle of microcanonical average, then the small system is distributed according to the canonical law
with a density\cite{khinchin1,darrigol}
\begin{equation}
\rho=\frac{1}{Z}e^{-\beta H}
\end{equation}
where $H$ is the hamiltonian of the small system and where the unique remain of the bath is the inverse temperature $\beta$. 
To arrive at the canonical, much more tractable, description, one can also remark that since the sole role of the bath is to fix the temperature of the system, it doesn't matter of what it is actually composed.
So that, one can perfectly imagine a huge (infinite) collection of components all identical in nature to the system under consideration, free to exchange energy. One is naturally led to the canonical law and to the concept of ensemble (at least in the canonical case).\cite{gibbs} 
From here, we will not continue further in the development of the logical consequences of the canonical 
distribution. The ambition of this first part was to clarify some aspects of equilibrium statistical mechanics.

\section{Steps toward non-equilibrium}
As it is well known and strongly emphasized in the specialized literature, the non-equilibrium situation is not as developed as the equilibrium case is. In particular, most physicists will agree to say that one cannot speak of a non-equilibrium statistical mechanics, in sharp contrast with equilibrium statistical theory. Nevertheless, in recent years there have been several developments that have led to general results not restricted to the vicinity of the equilibrium regime (as linear response theory is). In this section we briefly present some aspects of that results, namely the fluctuation theorem and the Jarzynski equality.

\subsection{Jarzynski equality}
The Jarzinsky equality is an ``unexpected'' equality relating equilibrium quantities with the average of a non equilibrium process that can be very far from equilibrium.\cite{jarzynski1}
More specifically, if one takes a system that is initially in a state of equilibrium at inverse temperature $\beta$ with an external (work) parameter denoted by $A$ and then if, within a finite time (so that we drive the system out of equilibrium), one tunes the external work parameter to the new value $B$, we will perform on the system some amount of work $W$ which is specific of the actual microstate of the system. If we repeat this experiment many times and record the values $W$ then the Jarzynski equality states that
\begin{equation}
\langle e^{-\beta W}\rangle \equiv \int {\rm d}W\; \rho(W) e^{-\beta W}=e^{-\beta \Delta F}
\label{jar}
\end{equation}
where $\rho(W)$ is the distribution of work within the given protocol (how we have tuned the work parameter) and $\Delta F= F_B-F_A$ is the free energy difference between the equilibrium states at temperature $\beta^{-1}$ with respectively external parameters $B$ and $A$.
In order to demonstrate the Jarzynski equality, we will follow the lines developed in ref.\cite{jarzynski2}, and present the case of a classical Hamiltonian dynamical system. But it is necessary to mention here that such an equality was also derived in the quantum case\cite{yukawa,mukamel,monnai} and within a stochastic markovian dynamics too\cite{jarzynski1,crooks1}.
The Jarzynski equality was soon tested experimentally. One can see ref.\cite{ritort} for a recent review.

The starting point of the demonstration given in ref.\cite{jarzynski2} is to consider the hamiltonian of the system and its thermal environment, together with an interaction term:
\begin{equation}
{\cal H}(\Gamma;\lambda)={\cal H}_s(x,\lambda)+{\cal H}_e(y)+
{\cal H}_{int}(x,y)
\end{equation}
where the subscript $s$($e$) refers to the system(environment) hamiltonian with collective dynamical variables represented by $x$($y$). $\Gamma=(x,y)$ is the phase space coordinate of the full system plus environment dynamical system.  
The interaction term ${\cal H}_{int}(x,y)$ is supposed to be small enough in order that one can interpret
${\cal H}_s$ as the internal energy of the system of interest. 
Let us initially consider that the system and environment are described by the thermal equilibrium Gibbs state with work parameter $\lambda=A$
\begin{equation}
p(\Gamma)=\frac{1}{Z(A)}\exp\big(-\beta{\cal H}(\Gamma;A)\big) 
\label{gibbs}
\end{equation}
where $Z(A)$ is the normalization factor. 
When we vary the work parameter from the initial value $A$ to the final value $B$ within a time $\tau$ and a predefined protocol, the energy change of the system over the given microscopic trajectory $\{\Gamma\}_t$ is given by
\begin{eqnarray}
{\cal H}_s(x_{\tau};B)-{\cal H}_s(x_{0};A)=\int_0^{\tau}{\rm d}t\;
\dot\lambda \frac{\partial {\cal H}_s}{\partial \lambda}(x_t;\lambda_t)
\nonumber \\ 
 +
\int_0^\tau {\rm d}t\;
\dot{x} \frac{\partial {\cal H}_s}{\partial x}(x_t;\lambda_t)
\end{eqnarray}
where the first integral is interpreted as the work $W$ performed on the system and so the second integral is the heat absorbed during the process. One may notice that, since the dynamics is hamiltonian, the work $W$ is given by the total energy difference 
\begin{equation}
W={\cal H}(\Gamma_{\tau};B)-{\cal H}(\Gamma_{0};A)\; .
\end{equation}
To compute the average $\langle e^{-\beta W}\rangle$ over the initial Gibbs state, one can use for the work the previous expression and write
\begin{equation}
\langle e^{-\beta W}\rangle = \int {\rm d}\Gamma_0\;p(\Gamma_0)
e^{-\beta[{\cal H}(\Gamma_{\tau};B)-{\cal H}(\Gamma_{0};A)]}
\end{equation}
and using the expression (\ref{gibbs}), one arrives at
\begin{equation}
\langle e^{-\beta W}\rangle = \frac{1}{Z(A)} \int {\rm d}\Gamma_0 \;
e^{-\beta{\cal H}(\Gamma_{\tau};B)}\; .
\end{equation}
Now, by the canonical change of variables $\Gamma_0\rightarrow \Gamma_\tau$, and using Liouville theorem on the invariance of the measure under canonical transformations, one finally arrives at
\begin{equation}
\langle e^{-\beta W}\rangle = \frac{1}{Z(A)} \int {\rm d}\Gamma_\tau \;
e^{-\beta{\cal H}(\Gamma_{\tau};B)}\; =\frac{Z(B)}{Z(A)} .
\end{equation}
This last relation is looking like the Jarzynski equality (\ref{jar})
but the partition functions entering here are that of the full system and environment. Nevertheless, one can arrive at the Jarzynski equality
(\ref{jar}) simply by noting that if we are able to neglect the interaction term (which has to be small enough), then the partition functions factorize into an environment term independant of the work parameter and a system term depending on $\lambda$.  The  ratio $Z(B)/Z(A)$
can then be rewritten as the Jarzynski equality
\begin{equation}
\langle e^{-\beta W}\rangle =\frac{Z(B)}{Z(A)} \approx 
\frac{
\int {\rm d}x \;e^{-\beta{\cal H}_s(x;B)}
}
{
\int {\rm d}x \;e^{-\beta{\cal H}_s(x;A)}
}
=\frac{Z_s(B)}{Z_s(A)} =e^{-\beta \Delta F}
\end{equation}
where this last form referes only to quantities pertaining to the system of interest.

Let us somehow discuss on a physical ground the Jarzynski equality. Indeed, if we perform on the system a reversible process by varying slowly enough the work parameter $\lambda$ from $A$ to $B$, then it is clear that 
$$
\Delta F=\tilde{W}
$$
where $\tilde{W}$ is the thermodynamical work performed on the system. 
If the switching is faster enough such that the system has no time to equilibrate with the new work parameter, then the second law of thermodynamics states that the work $\tilde{W}$ has to be larger so that one has for a general process the least work principle:
\begin{equation}
\tilde{W}\ge \Delta F
\end{equation}
which can be written
\begin{equation}
\tilde{W}_{dis}=\tilde{W}-\Delta F \ge 0
\end{equation}
where the equality holds in the reversible case. 
If we rewrite the Jarzynski equality with $W_{dis}=W-\Delta F$, we have simply
\begin{equation}
\langle \exp\big(-\beta W_{dis}\big)\rangle =1\; .
\end{equation}
Using Jensen's relation $\langle \exp(x)\rangle \ge \exp (\langle x\rangle)$, we see that the Jarzynski equality implies
\begin{equation}
\langle W_{dis}\rangle \ge 0
\end{equation}
which is the least work principle if one identifies the average work 
$\langle W\rangle$ with the thermodynamical work $\tilde{W}$.
From this, roughly speaking, one sees that if $\sigma$ is a parameter that controls irreversibility (one may think of the entropy production), the density distribution $\rho_{\sigma}(W)$
$$
\lim_{\sigma\rightarrow 0}\rho_{\sigma}(W)=\delta(W-\Delta F)
$$
in the reversible case $\sigma=0$. 
Moreover, in the linear response regime, one expects gaussian fluctuations of the work $W$ such that the density takes the form
$$
\rho(W)=\frac{1}{\sqrt{2\pi \theta}}
\exp\left(-\frac{(W-W_0)^2}{2\theta}\right)
$$
near equilibrium, where $\theta=\langle (W-\langle W\rangle)^2\rangle$ is the variance of the distribution and $W_0=\langle W\rangle$ the mean value of the work. Using this expression together with the Jarzynski equality, one can relate the fluctuation of the work to the dissipated
work $\langle W_{dis}\rangle$, or to the entropy production if one notes that $T\sigma=\langle W_{dis}\rangle=T\Delta S-Q=\langle W\rangle-\Delta F$:
$$
\langle W_{dis}\rangle=\frac{1}{2}\beta \theta\; ,
$$
or equivalently
$$
\sigma= k_B \frac{1}{2}\langle(w-\langle w\rangle)^2\rangle
$$
with $w\equiv W/k_BT$. 
We recover here a fluctuation dissipation theorem, relating the dissipated work to the fluctuation of it.
Within this reduced work variable $w$, the density of the reduced  dissipated work $w_d$ is rewritten
$$
\pi_{\sigma}(w_d)=\frac{1}{\sqrt{2\pi \sigma}} \exp\left(-\frac{(w_d -\sigma/2)^2}{2\sigma}\right)
$$
where the Boltzmann constant $k_B$ has been absorbed into the definition of the entropy production. In the reversible limit, one has
$$
\lim_{\sigma\rightarrow 0}\pi_\sigma(w_d)=\delta(w_d)\; ,
$$
that is no dissipation at all.

The very interesting feature of the Jarzynski equality is that it provides a method to recover free-energy differences from non-equilibrium experiments. This has been done recently for exemple on single molecule stretching experiments\cite{ritort,hummer,liphart}. The possibility of recording free-energy differences was also experienced  on simple theoretical models as one can see \cite{mazonka,douarche,ritort2} for some examples and \cite{ritort,maes} for recent reviews on the experimental and theoretical aspects respectively.

\subsection{Fluctuation theorem}
Soon after the derivation of the Jarzynski relation it was demonstrated\cite{crooks2} that such result can be derived from a much more general theorem, namely the fluctuation relation which is a statement on the fluctuations of the entropy production. The first example of such a relation was obtained numerically in ref.\cite{evans} for steady states systems and was soon after derived for driven thermostated deterministic systems\cite{evans2}, thermostated steady state systems with deterministic\cite{gallavotti2} and stochastic dynamics\cite{kurchan,lebowitz,maes2}. The theorem can be written in the following form:
\begin{equation}
\frac{P(\sigma_{\tau})}{P(-\sigma_\tau)}\simeq\exp(\tau \sigma_\tau)
\end{equation}
where $P(\sigma_\tau)$ is the probability of having the entropy production rate $\sigma_\tau$ over a time $\tau$. 
This relation is in fact an asymptotic statement which is valid in the $\tau\rightarrow \infty$ limit. The precise statement of the theorem could be enonced in the following way:\cite{gallavotti,maes} 
Let $\sigma(\Gamma)$ be the entropy production rate in the stationary non-equilibrium state of the dynamical system represented at the initial time by the phase space point $\Gamma$. Let us define the ``dimensionless average entropy creation rate'' 
$$
p=\frac{1}{\sigma_+ \tau}\int_{0}^{\tau}{\rm d}t\; \sigma(\Gamma_t)
$$
where $\sigma_+\ge 0$ is the average entropy creation rate in the steady state and where the $+$ subscript emphasizes the positivity of it.
The fluctuation theorem states that 
\begin{equation}
\zeta(p)=\zeta(-p)+p\sigma_+\; ,
\end{equation}
where $\zeta(p)$ is the large deviation function of $p$ defined as
$$
\zeta(p)=\lim_{\tau\rightarrow \infty} \frac{1}{\tau} \ln \pi_\tau(p)
$$
with $\pi_{\tau}(p)$ the probability distribution of the variable $p$.
The validity of the fluctuation theorem is very broad since it was proved within several different dynamical contexts, such as reversible hyperbolic dynamical systems \cite{gallavotti2} or nondeterministic dynamics \cite{kurchan,lebowitz,evans2}. 
It was proved also that the fluctuation theorem in the limit of a vanishing entropy production rate implies the linear (near-equilibrium) fluctuation-dissipation theorem \cite{gallavotti3,gallavotti4,kurchan}.
The theorem was also tested in a number of numerical investigations \cite{bonetto} and recently on granular materials and turbulent flows \cite{ciliberto}. We may mention also some recent extension for stochastic systems which are not able to equilibrate with their environment in the limit of zero external drive \cite{zamponi}.

Whereas the fluctuation theorem is an asymptotic relation, Crooks derived a very interesting identity in the case of stochastic microscopically reversible dynamics \cite{crooks1,crooks2}. This identity reads
\begin{equation}
\frac{P_F(\Omega)}{P_R(-\Omega)}=e^\Omega
\end{equation}
where $\Omega$ is the entropy production of the system driven for some time $\tau$ within a forward protocol $\lambda_F(t)$ to a new state. $P_F$ is the distribution of entropy production in the forward process, while $P_R$ is the distribution of the entropy production in the backward or reverse process, that is when the system is driven in a time-reversed manner.

A particularly intersting case is one in which the system is initially in contact with a thermal heat bath. The forward protocol is such that   the system being initially (by convention we will take the far past $t=-\infty$ to label the initial time) at equilibrium with an external parameter $\lambda_-$, is driven out of equilibrium by varying the control parameter to a new value $\lambda_+$. The variation of $\lambda$ takes a finite time $\tau$, let us say within the interval $[0,\tau]$. For times larger than $\tau$, the system is equilibrating with the heat bath and terminates in an equilibrium state characterized by the new external parameter value $\lambda_+$.
Since the equilibrium entropy is given by $S=\langle -\ln \rho(\Gamma)\rangle=-\sum_\Gamma \rho(\Gamma)\ln \rho(\Gamma)$, it is quite natural to identify 
$$
s(\Gamma)=-\ln \rho(\Gamma)
$$
as the entropy of the microstate $\Gamma$. The entropy production for a given trajectory in phase space is then 
\begin{equation}
\Omega=\ln \rho(\Gamma_{-\infty})-
\ln \rho(\Gamma_{\infty})-\beta Q
\end{equation}
where $\rho(\Gamma_{\pm \infty})=\rho(\Gamma,\lambda_\pm)$ are the canonical equilibrium densities associated to the initial and final equilibrium states and where $Q$ is the heat exchanged during the process with the heat bath.
Introducing into the previous expression the explicit equilibrium density 
$$
\rho (\Gamma,\lambda)=e^{\beta F(\lambda)-\beta E(\Gamma,\lambda)}\; ,
$$
one arrives, with $\Delta E= W+Q$, for the entropy production at
$$
\Omega=\beta  [W-\Delta F]\; .
$$
So that the Crooks fluctuation  identity can be written as a relation for the work performed on the system:
\begin{equation}
\frac{P_F(W)}{P_R(-W)}=e^{\beta(W-\Delta F)}\; .
\end{equation}
From this last expression, one can see that the Jarzynski identy is trivially recovered since
\begin{eqnarray}
\langle e^{-\beta W}\rangle=\int P_F(W)e^{-\beta W}\; {\rm d}W=
\nonumber \\
e^{-\beta \Delta F}\int P_R(W)\; {\rm d}W=e^{-\beta \Delta F}\; 
\end{eqnarray}
where the normalisation condition of $P_R$ has been used. 

An interesting point is when one considers a cyclic transformation. In this case the final state is identical to the initial state and $\Delta F=0$ so that the entropy production is given by $-Q/T$. Utilising again Jensen's inequality we obtain that in average the entropy production is positive, which means that since work is performed on the system, heat is tranfered to the bath (and not the converse which is Clausius basic statement). 

\section{Work on the $2d$-Ising model}
In this paper, we discussed aspects of equilibrium statistical mechanics and focused some attention on non-equilibrium fluctuation relations. 
In order to give a concret example in the out of equilibrium situation, we give some preleminary results for work distributions obtained on the $2d$-Ising model \cite{chatelain}. 
The zero field Hamiltonian is given by 
$$
{\cal H}= -\sum_{(ij)} S_iS_j
$$ 
where the $S=\pm 1$ are classical Ising variables and where the sum extends over nearest neighbors. In the numerical simulations, we use a Metropolis dynamics. The protocol we use is such that at the intial time $t=0$, the system beeing in an equilibrium state at inverse temperature $\beta$ and zero  field, we swich on the field for a time $\tau$ with
a linear law:
$$
h(t)=\dot{H} t = \frac{H}{\tau} t\;.
$$
For each initial realization, we compute the work W performed on the system by the external field with 
$$
W=-\dot{H}\int_0^\tau {\rm d} t\; M(t) 
$$
where $M(t)=\sum_i S_i(t)$ is the instantaneous magnetization of the system. 

\begin{figure}[ht]
  \centering
  \begin{minipage}{\textwidth}
  \epsfig{file=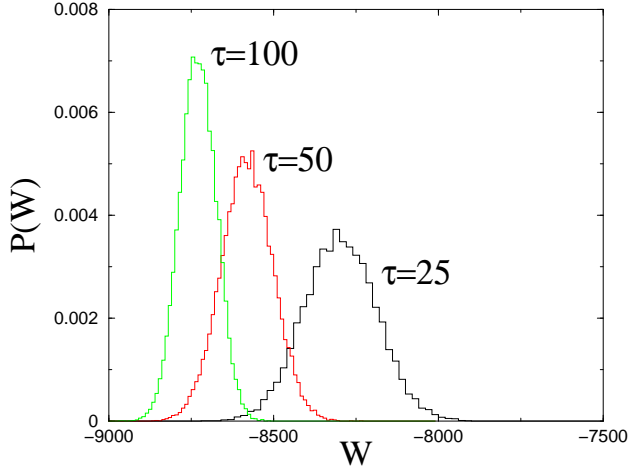,width=0.55\textwidth}
  \end{minipage}
  \caption{Work probablility density distribution of the 2d Ising model  at $\beta=0.2$ (paramagnetic initial phase) with a linear varying field from $\beta h(0)=0$ to $\beta h(\tau)=0.3$ for several time intervals $\tau$. The system size is $L\times L=128^2$.
	}
\label{fig1}
\end{figure}

We first show on figure~1 the work distribution obtained with an initial paramagnetic state for three different protocols with ending field $h=1.5$ reached after time $\tau=25$, $\tau=50$ and $\tau=100$, that is 
smoother and smoother protocols. 
As it can be seen on figure~1, the distributions are gaussian and behave as predicted in the previous section, with a dissipated work proportional to the variance of the distribution. From the shift to the left (negative value), one can extract the free energy difference, since as $\tau$ is increasing, the transformation is less and less irreversible (the entropy production is smaller and smaller). 
Since the distribution is gaussian (near equilibrium like regime), the free energy difference is given by
$$
\Delta F= \langle W \rangle -\frac{\beta}{2} \theta
$$
where $\theta$ is the variance of the distribution. We have verified here that the Jarzynski equality holds, since the previous relation is satisfied and gives the same value  $\Delta F$ for different $\tau$ values.
Moreover, from our data we find a linear dependence of the dissipated work with the perturbing speed $\dot{H}$, as it is expected near equilibrium or more generally in the gaussian case \cite{ritort}, with
$$
\langle W_{dis}\rangle \simeq \alpha(H) L^2 \dot{H} 
$$
with a field dependent slope $\alpha(H)$. From the analyse of the work distributions for several ending fields, we arrived numerically at 
an expression for $\alpha(H)$ which is very well fitted by:
$$
\alpha(H)=\gamma \mu(H)=\gamma \mu \tanh (\beta \mu H)
$$
with $\gamma\simeq 0.43$ and $\mu(\beta=0.2)\simeq 3.65$,
where $\mu(H)$ is the paramagnetic magnetization. 
Indeed, if one extracts the free energy difference as a function of the field $H$, one obtains a very good fit with the paramagnetic solution 
$$
\Delta F(H)=-\frac{L^2}{\kappa \beta} \left[\ln \frac{\sinh (2\beta\mu H)}{\sinh(\beta\mu H)}-\ln2\right]
$$
with a magnetic moment $\mu\simeq 3.65$ and volume $\kappa\simeq 4.6$.
If one uses the equilibrium magnetization extracted from that expression: 
$$
M(H,\beta)=-\frac{\partial F}{\partial H}=\frac{L^2}{\kappa }\mu \tanh(\beta\mu H)\; ,
$$ 
the dissipated work can be expressed as
$$
\langle W_{dis}\rangle \simeq 2 \dot{H} M(H,\beta)
$$
so that it depends only on the equilibrium quantities of the paramagnetic gaz. 

If we decrease the temperature of the initial equilibrium state, at a moment we will arrive at the critical point. 
Of course, since our system is finite, the correlation length will not diverge since it will reach the boundaries of the system. Due to the long range critical correlations, one expects a different behaviour of the work performed on the system along the same linear protocol as in the paramagnetic phase. Indeed, it is the case as it can be seen on figure~2 where the work distribution is plotted for two ending fields, using three different protocols with $\tau=25$, $\tau=50$ and $\tau=100$. Insteed of having one pronouced peak as in the paramagnetic case, we have to face with two peaks distributions. 

\begin{figure}[ht]
  \centering
  \begin{minipage}{\textwidth}
  \epsfig{file=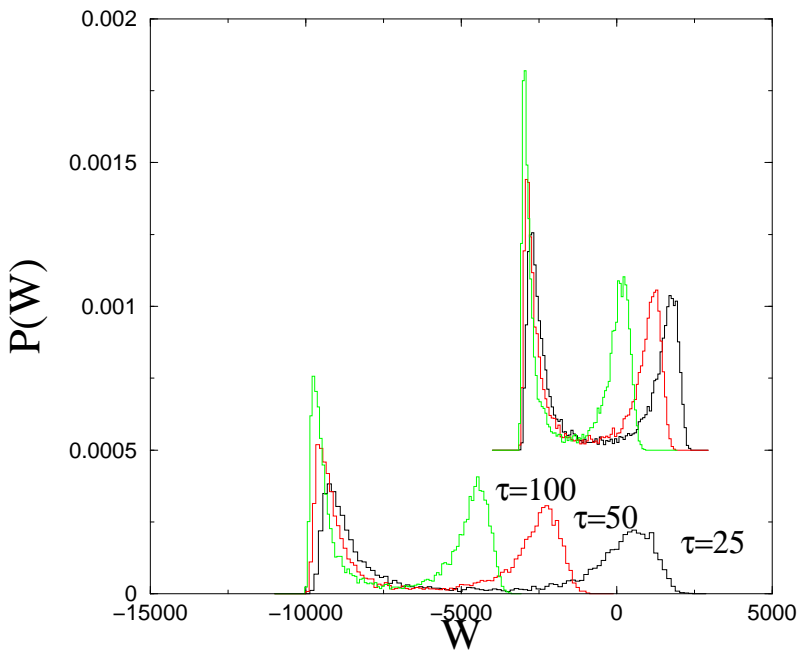,width=0.70\textwidth}
  \end{minipage}
\caption{ Work probablility density distribution of the 2d Ising model  at $\beta_c$ (critical initial phase) with ending fields $\beta_c h=0.1$ and $\beta_c h=0.3$ for several time intervals $\tau$. The system size is $L\times L=128^2$. The distribution with $\beta_c h=0.1$ has been shifted vertically by an amount of $5.10^{-4}$.}
\label{fig2}
\end{figure}

What we see is that the left most peak, sitting near by the reversible work $\Delta F$, is moving very slitly toward negative values for larger time intervals $\tau$. The right most peak has a strong dependance with the perturbing speed
$\dot{H}$ as it is obvious from figure~2 and seems to converge asymptotically toward the reversible value $\Delta F$.
We can understand the structure of the distribution in the following way: since our system is finite, even at the critical temperature $\beta_c^{-1}$, there is a finite magnetization of the order $L^{d-\beta/\nu}$. When turning on the field, if the initial magnetization is pointing in the direction of the field, we expect a negative work of the order $-H M$, where $M$ is the initial magnetization. On the contrary, if the initial magnetization is pointing in the opposite direction, at least for very fast protocols, one will have a positif work $+HM$ since the system will not have enough time to react to the variation of the field. For lower field speed $\dot{H}$, during the time interval $\tau$, part of the magnetization will flip in the field direction, leading to a smaller work value. 
It is clear that the real situation is more complicated than the rougth description just sketched above. In fact, one has to take into account the domain growth, since it will be the leading process for the variation of the total magnetization, and so for the work performed on the system. 
This study is currently under investigation.

Let us finally discuss the ferromagnetic situation. In figure~3, we have presented the work distribution, for the three field speeds $\dot{H}$, obtained with an ending field $\beta h=0.1$, with $\beta=0.7$. What is seen is two very well separeted peaks, one sitting at a negative value $W_-$ and the other at a positive value $W_+$ of the same order of magnitude. Moreover, each peak has the same weigth $1/2$. For clarity, the right most peak has been plotted in an insert.  
\begin{figure}[ht]

  \centering
  \begin{minipage}{\textwidth}
  \epsfig{file=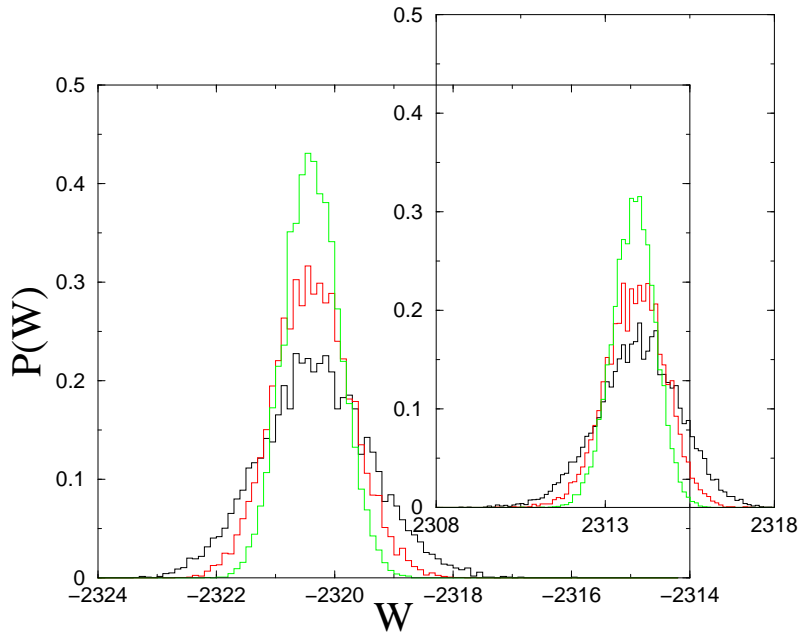 ,width=0.70\textwidth}
  \end{minipage}
\caption{ Work probablility density distribution of the 2d Ising model  at $\beta=0.7$ (ferromagnetic initial phase) with ending field $\beta h=0.1$ for several time intervals $\tau$. The system size is $L\times L=128^2$. }
\label{fig3}
\end{figure}
The peaks are more or less gaussian like (in fact the right peak is a little bit asymmetric and so deviates from the gaussian behaviour) and one can see that as the process gets slower, the peaks gets sharper. 
What is also manifest is the fact that the left peak is almost no more translated to the left as in the paramagnetic case. Indeed, it corresponds to the initial equilibrium magnetization pointing in the direction of the field, and since at $\beta=0.7$, the magnetization is nearly saturated, there is no magnetization gain to be expected during the swiching of the field. The work is then of the order of $-|HM_{eq}(\beta,0)|$. The right peak corresponds to a microstate with an initial magnetization pointing in the opposite direction of the field. In this case, if the field is not strong enough, the macroscopic magnetization will stay unalterated by the presence of the opposite field and we will have more or less $W_+\sim +|HM_{eq}(\beta,0)|$. In the slow swiching limit, the work distribution takes the form
$$
P(W)=\frac{1}{2} \delta_{W_-}+\frac{1}{2} \delta_{W_+}
$$
and so the average work $\langle W\rangle$ is given by
$$
\langle W\rangle=\frac{W_-+W_+}{2}
$$
while, using Jarzynski equality, the free energy difference $\Delta F$ is such that
$$
e^{-\beta\Delta F}\simeq \frac{1}{2} e^{-\beta W_-}
$$
where the argument of the exponential which is positive dominates largely the other exponential. From that, we have
$$
\Delta F \simeq -|W_-|+\frac{1}{\beta}\ln 2 \simeq -|W_-|
$$
where $\ln 2/\beta$ has been neglected since the work $W_-$ is extensive. Together with the previous expression, we have for the dissipated work
$$
\langle W_{dis}\rangle= \langle W\rangle-\Delta F\simeq \frac{W_+-W_-}{2}\simeq -\Delta F
$$
that is, the dissipated work is of the order of the reversible work (in magnitude). This striking result is due to the fact that almost all the irreversibility is putted in by the reversed magnetized microstates. If one uses as a macrostate only those microstates with the magnetization pointing in the field direction, the situation would have been more usual, that is gaussian with small dissipation. From the data presented here, the ratio of the dissipated work with the free energy is less than $5.10^{-4}$, which explains why there is no apparent shift toward the left of the curves for decreasing perturbation speed $\dot{H}$.

We close here the discussion of these numerical preliminary results. The point was to give an example where we can extract some equilibrium quantities within non-equilibrium (computer) experiments. However, one may notice that in order to obtain accurate results for the free energy difference from the Jarzynski equality, one has to sample out very efficiently the tails (at least in the left most part) of the distribution since we need to compute exponentials
of extensive quantities (work). But nevertheless, this program seems to be realistic enough and is currently under progress. Finally, 
one may think of realizing such situations experimentally, measuring nanoparticule's magnetization within a SQUID experiment \cite{dragi}. 

\section*{Acknowledgments}
Arnaldo Donoso, Bertrando Berches, Ricardo Paredes,  and all the people who made Mochima spring/summer school possible are all greatfully acknowledged. I would like to give a special Thank to the students who made the school to be so pleasant.
Part of the material presented here was discussed and done in collaboration with Christophe Chatelain. Je remercie le $\cal G$roupe $\cal M$ pour son soutien.


\begin{thebibliography}{99}
\bibitem{dragi} In principle such experiment is possible in our laboratory. The challenge is to persuade colleagues that it is indeed interesting enough.
\end{thebibliography}
\end{document}